\documentclass[manuscript, screen, nonacm]{acmart}
\usepackage[table]{xcolor}
\usepackage{tabularx}

\AtBeginDocument{%
  }

\setcopyright{acmlicensed}
\copyrightyear{2018}
\acmYear{2018}
\acmDOI{XXXXXXX.XXXXXXX}
\acmConference[CARS@RecSys '26]{Workshop on Context-Aware Recommender Systems, co-located with the 20th ACM Conference on Recommender Systems (RecSys 2026)}{September 28--October 2, 2026}{Minneapolis, MN, USA}
\acmISBN{978-1-4503-XXXX-X/2018/06}

\begin{document}

\title{Seeing the Context: Enhancing Recommender Systems with Image-Derived Contextual Signals}

\author{Tal Cordova}
\affiliation{%
  \institution{Coller School of Management, Tel Aviv University}
  \city{Tel Aviv}
  \country{Israel}}
\email{talcordova@mail.tau.ac.il}

\author{Tomer Geva}
\affiliation{%
  \institution{Coller School of Management, Tel Aviv University}
  \city{Tel Aviv}
  \country{Israel}}
\email{tomergev@tauex.tau.ac.il}

\author{Moshe Unger}
\affiliation{%
  \institution{Coller School of Management, Tel Aviv University}
  \city{Tel Aviv}
  \country{Israel}}
\email{mosheunger@tauex.tau.ac.il}

\renewcommand{\shortauthors}{Cordova et al.}

\begin{abstract}
    Contextual information, capturing the circumstances of a user-item interaction, is central to recommender systems. Prior work draws context from location, time, or reviews, but not images; multimodal recommender systems mainly use images to enrich item or user representations, not identify situational context. We propose a new representation of context derived from images, spanning physical, social, and modal categories learned via a vision-language model. We introduce ICE-Fuse, a pipeline for evaluating this representation that fuses these categories and integrates them into a context-aware recommender system, using TripAdvisor data and Review-aware Graph Contrastive Learning as the recommendation algorithm. Image context does not outperform established signals standalone, but improves them combined, indicating complementary information. Semantic analysis shows image- and review-derived context capture distinct aspects of the interaction, positioning images as complementary context.
\end{abstract}


\maketitle

\section{Introduction}
Contextual information, which is defined as the circumstances in which a user consumes an  item, plays an important role in recommender systems (RS) \cite{adomavicius2022cars}. Context-aware recommender systems (CARS) typically acquire this information from location, time, or weather \cite{adomavicius2022cars}, or from textual reviews \cite{bauman2022knowthycontext}. While recommender systems use images extensively \cite{liu2024survey}, they do so to enrich item or user representations \cite{jeong2024multimodal}, rather than to infer situational context. This is important since images are captured during the experience itself and can preserve contextual details such as the formality (e.g., casual attire versus a set table with candles) or ambience (e.g., a lively, crowded room versus a quiet, intimate one), that are not detailed in other contextual sources. Furthermore, context from textual reviews may lose these details since textual reviews are written after the fact and shaped by retrospective memory \cite{fredrickson1993peakend}. In this work, we propose acquiring contextual information from images to improve recommender systems. Specifically, we define three types of image-based context categories: physical, social and modal context \cite{adomavicius2011cars}. These categories can be inferred from the image using visual cues, for example, the number of plates can be an indication of a social gathering.

Our research questions are the following: 

\textbf{RQ1}: How does context acquired from images affect recommendation performance compared with established contextual signals? 

\textbf{RQ2}: Does image-based context improve recommendation performance when combined with other contextual and content signals? 

To answer these questions, we introduce ICE-Fuse, a pipeline for evaluating the value of different types of contextual features in RS. For each context category (i.e., physical, social, and modal), we prompt a vision language model (VLM) to return a contextual description for each picture, and then fuse those separate descriptions using a gating mechanism for a unified context representation. The context is then integrated into a well-known recommendation model.

In this paper, we make the following contributions. First, we propose, for the first time, a new contextual signal from images to recommendation systems. Second, we introduce a designated evaluation pipeline, where we evaluate this signal compared to other signals, or as a complementary signal, in terms of recommendation performance in rating prediction tasks. Third, we identify several contextual features that are distinct to images, and are not present in other contextual sources. Our results show that image context does not outperform other context signals on its own in terms of rating prediction performance, but it could improve predictive performance when integrated with other context signals, showing that image context carries complementary contextual information to existing sources.

\section{Method}

Figure 1 illustrates the \textbf{ICE-Fuse} pipeline: context descriptions for each interaction are fused into a single image-context representation and incorporated into Review-aware Graph Contrastive Learning (RGCL) \cite{shuai2022rgcl}, a graph-based, review-aware recommendation model that we modified to integrate context during training, allowing image-derived information to refine the learned user and item representations. These representations are then used for rating prediction without requiring images at inference time. 

ICE-Fuse follows the CARS taxonomy, in which context is represented by three categories: physical (e.g., setting, ambiance, time of day), social (e.g., group size, occasion), and modal (e.g., formality, mood)  \cite{adomavicius2011cars}.  For each category, a dedicated prompt instructs the VLM to identify relevant situational cues and return a textual context description, avoiding item-content information (e.g., specific dishes or restaurant identity) unless it evidences situational context. For each image associated with interaction (u, i), the VLM produces one description per category, which is encoded and projected via a category-specific trainable MLP: 

\begin{align}
\mathbf{c}_{ui}^{k}
    &= E_{\text{text}}\!\left(d_{ui}^{k}\right), \qquad
\mathbf{h}_{ui}^{k}=\operatorname{MLP}_{k}\!\left(\mathbf{c}_{ui}^{k}\right),
\qquad
\mathcal{K}=\{\text{physical},\text{social},\text{modal}\}.
\end{align}

Here, $d_{ui}^{k}$ is the VLM-generated description of category $k$, $\mathbf{c}_{ui}^{k}$  its text-embedding vector, and $\mathbf{h}_{ui}^{k}$ its projected representation, dimensioned to match the user/item embedding size. Only projected representations pass to fusion. 

\begin{figure}
  \centering
 \includegraphics[width=0.75\linewidth]{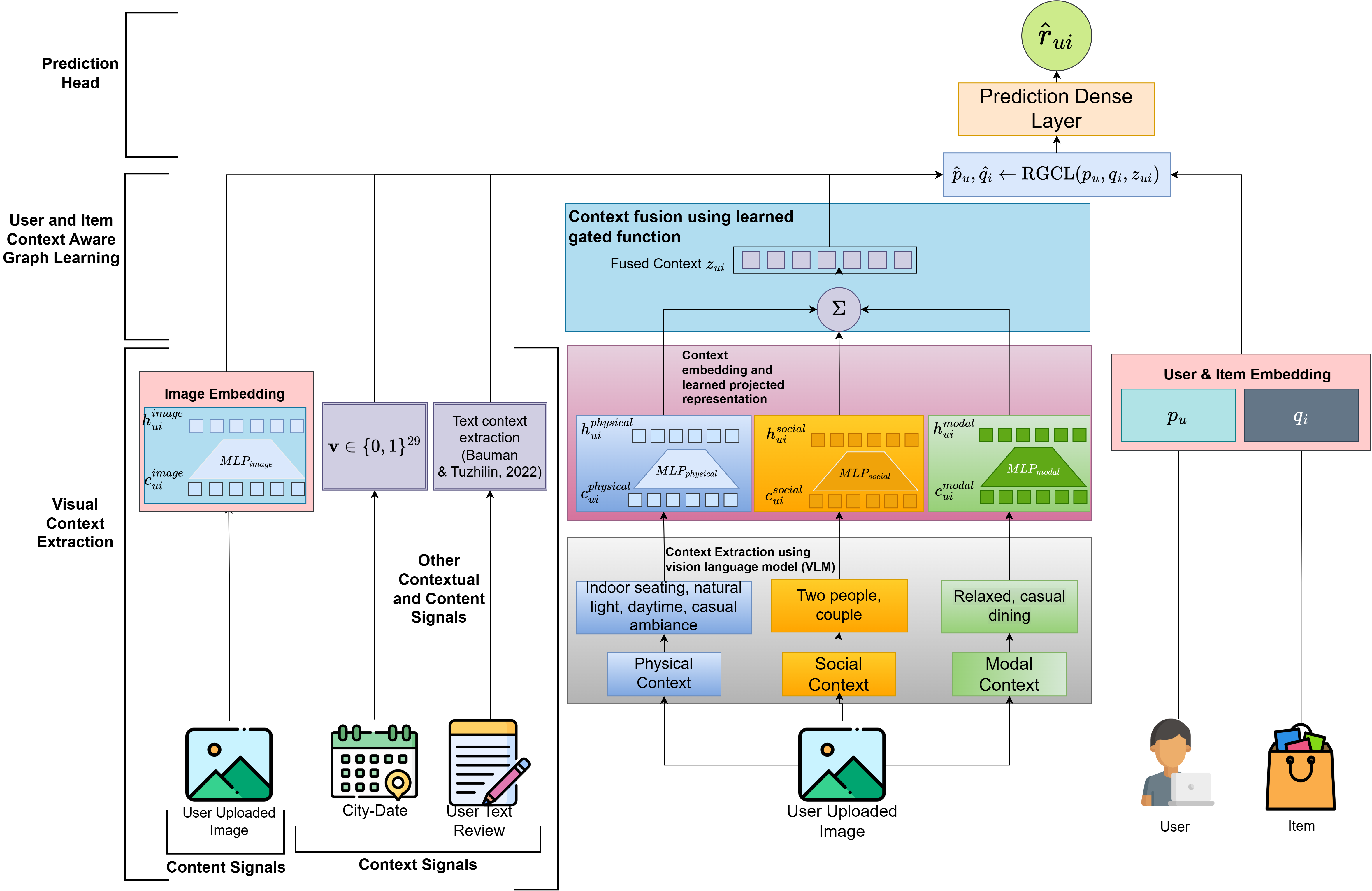}
  \caption{Overview of ICE-Fuse, the proposed image-context extraction and signal-fusion framework for CARS.
  }
  \label{fig:Method}
\end{figure}

\subsection{Context Fusion, Integration and Rating Prediction}
\label{sec:contextAcquisition}

Rather than using fixed weights, ICE-Fuse learns each category's contribution via a gating network:

\begin{align}
\boldsymbol{\gamma}_{ui}
    &= \operatorname{softmax}\!\left(
       g(\mathbf{p}_u,\mathbf{q}_i)
       \right), \quad
\mathbf{z}_{ui}= \sum_{k \in \mathcal{K}}
       \gamma_{ui}^{k}\mathbf{h}_{ui}^{k},
\qquad
\mathcal{K}=\{\text{physical},\text{social},\text{modal}\}.
\end{align}

Here, $g$ is a learned gating network that produces one score per context category, transformed by softmax into weights $\gamma_{ui}$ that sum to one, so that the relative contributions of physical, social, and modal context can vary across user-item interactions rather than remaining fixed across the dataset. The resulting vector $z_{ui}$ represents the fused image context for interaction $(u, i)$, and serves as the contextual representation of the review edge between user  and item  within the RS model. The chosen RS model is RGCL \cite{shuai2022rgcl}.

RGCL was chosen to address an inference-time constraint: since images are unavailable at inference, the recommendation model must integrate contextual features into user and item representations during training. RGCL supports the integration of review information, and we modified it to incorporate contextual information. Context acquisition and fusion occur during training; the resulting representations $(\hat{p_u}, \hat{q_i}$) are used directly for prediction. Rating prediction is performed by a lightweight fully connected network that operates only on these representations, first constructing a joint interaction representation and then mapping it to a scalar rating.

\subsection{Alternative Context and Content Signals}
\label{sec:alternativeSignals}

To evaluate image context, we compare it with other signals. The first baseline is without any context (\textit{no-context}). This baseline is similar to a graph-based CF \cite{he2020lightgcn} model with the node-discrimination (ND) term of RGCL present \cite{shuai2022rgcl}. In addition, we compare with existing contextual signals. 

We consider two \textbf{contextual signals}: \textit{city-date} (29 one-hot features from restaurant city, weekend status, day of week, month, and season) \cite{adomavicius2011cars} and \textit{text-based context} \cite{bauman2022knowthycontext}, using 52 expert-identified phrases mapped to nine features (Table~\ref{tab:context-phrases}). 

Lastly, we also compare with one content signal, a direct \textit{image embedding}. Each is projected via a signal-specific MLP into the shared latent space, enabling standalone comparison with image context and, in combined configurations, concatenation with it. The image embedding uses the same review image as the image-context signal but encodes it directly via CLIP without VLM-based interpretation.

\begin{table}[h]
  \caption{Contextual features and associated phrases used in the text-based context baseline}
  \label{tab:context-phrases}
  \begin{tabular}{ll}
    \toprule
    Feature Name&Phrases\\
    \midrule
    Company&Friends, brother, family, husband, couples, work, tourist, locals, 2 people, 4 people, 5 people\\
    Occasion&Celebrate, birthday, occasions, several occasions, Christmas \\
    Type of meal&Having dinner, had dinner, lunch, breakfast, brunch, dining \\
    Time of day&Afternoon, evening, opening \\
    Day of week&Saturday, Saturday night, Monday, weekday\\
    Time of year&Christmas, this year \\
    Reservation&Had reservation, have reservation, made reservation, reservation, booked, reserved, had wait\\
    Prior visits&Three times, 3 times, visited, returned, discovered, found restaurant, found \\
    Advice&Reading reviews, heard, advised, recommendation, recommendations, read, TripAdvisor \\
  \bottomrule
\end{tabular}
\end{table}

\section{Evaluation}
\label{sec:evaluation}

We use ICE-Fuse to evaluate the recommendation performance of image context on a public TripAdvisor restaurant-review dataset \footnote{\url{https://zenodo.org/records/5644892}} spanning five European cities (Madrid, Barcelona, Moscow, Istanbul, Gijón), using reviews from 2010–2019 and retaining users and restaurants with at least seven image reviews. After filtering, we were left with 39,956 reviews, 1,441 users, and 1,341 items. Of the 39,956 total reviews, 22,220 contain images. Not all reviews contain images because we sample all reviews from qualifying users, not only those with images.

Image context is acquired using Gemma 3 12B IT \footnote{\url{https://huggingface.co/google/gemma-3-12b-it}} \cite{gemmateam2024} via Ollama (temperature 0.5, max 2048 tokens, thinking disabled), with descriptions encoded via e5-small-v2 (384-dim) \footnote{\url{https://huggingface.co/intfloat/e5-small-v2}} \cite{wang2024e5}. Direct image embeddings use CLIP ViT-B/32 (512-dim) \footnote{\url{https://huggingface.co/openai/clip-vit-base-patch32}} \cite{radford2021clip}. 

All contextual and content representations are projected to 64 dimensions, matching the user/item embedding size. We use a global time-based split, training on historical interactions from 2010–2016 and testing on 2017–2019. The train set contains 18,202 samples, of which 10\% were chosen randomly as the validation set. The test set contains 21,754 samples. Models are trained with Adam (lr = 0.001) and early stopping (patience 25) on validation loss, combining rating-prediction MSE with RGCL's contrastive losses ($\alpha$ = $\beta$ = 0.5). All experiments are conducted on a single NVIDIA RTX 6000 Ada GPU.

\section{Results}
\label{sec:results}

We use ICE-Fuse to evaluate the recommendation performance of image context along three dimensions. We first address RQ1 and RQ2 by comparing image context against established contextual and content signals, both standalone and in combination. We then examine semantic overlap between image- and review-derived context, followed by the gating mechanism's internal weighting across the physical, social, and modal categories. Statistical significance is assessed using paired two-sided Wilcoxon signed-rank tests \cite{demsar2006statistical}.

\subsection{Recommendation Performance}
\label{sec:recpPerformance}

Table~\ref{tab:context-signals} reports standalone rating-prediction performance. Image context significantly outperforms the no-context baseline, confirming that it carries recommendation-relevant signal. However, it does not outperform city-date or text-based context as a standalone predictor; addressing RQ1, image context does not exceed established contextual signals in isolation. We therefore examine whether image context adds value when combined with these signals. Table~\ref{tab:context-enhancement} addresses RQ2 by combining image context with each established signal. Adding image-based context to city-date context or to image-embedding reduces RMSE significantly compared to those signals alone. Adding image-based context to text-based context shows improvement compared to the signal on its own, though this difference is not statistically significant. Across all three signals, adding image context improves performance, indicating that image context contributes situational information distinct from existing contextual and content signals, even where the magnitude of that contribution varies by signal. 

\begin{table}[h]
\caption{Standalone rating-prediction performance of the evaluated contextual signals.}
\label{tab:context-signals}
\renewcommand{\arraystretch}{0.9}
\begin{tabular}{l p{4cm} l}
\toprule
Context Type & Description & RMSE \\
\midrule
No context & CF signal, no context & 0.954 \\
City date & Explicit context from city and timestamp & 0.903 \\
Text based context & Context acquired from reviews \cite{bauman2022knowthycontext} & \textbf{0.899} \\
Image based context & Proposed new context signal & 0.910 \\
\bottomrule
\end{tabular}
\end{table}

\begin{table}[h]
\caption{ Rating-prediction performance before and after combination with image context. Significance is reported relative to the strongest alternative to image context using a paired two-sided Wilcoxon signed-rank test. *** $p$ < 0.001, ** $p$ < 0.01, * $p$ < 0.05.}
\label{tab:context-enhancement}
\renewcommand{\arraystretch}{0.9}
\begin{tabular}{llll}
\toprule
Context Type & City-Date Context & Text-Based Context & Image Embedding \\
\midrule
Contextual Signal & 0.903 & 0.899 & 0.908  \\
Contextual signal combined with image context & \textbf{0.902***} & \textbf{0.896} & \textbf{0.903***} \\
\bottomrule
\end{tabular}
\end{table}

\subsection{Semantic Complementarity of Image and Review Context}
\label{sec:semnticComp}

\begin{figure}
  \centering
 \includegraphics[width=0.75\linewidth]{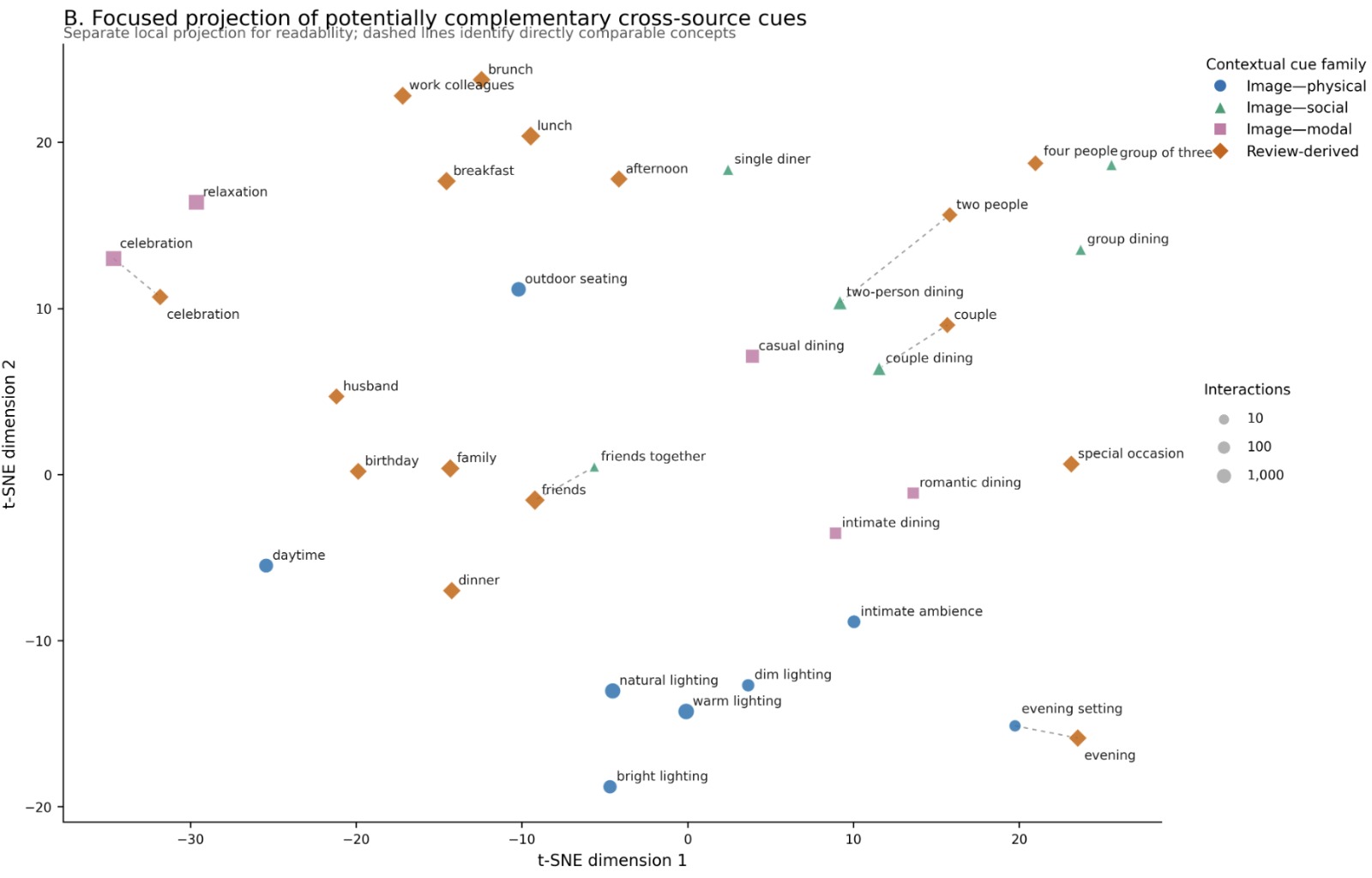}
  \caption{Common phrases and context in text based context and image based context.
  }
  \label{fig:Results}
\end{figure}

To examine whether image-derived and review-derived context capture distinct situational information, we project embeddings of both sources into a shared space using t-SNE and inspect their proximity (Figure~\ref{fig:Results}). A small number of concepts appear in close proximity across sources, for example, "two people" (review) and "two-person dining" (image), or "evening" (review) and "evening setting" (image), indicating some direct semantic overlap. The majority of concepts, however, remain source-specific: image-derived context uniquely captures fine-grained ambience (e.g., natural, dim, and warm lighting; outdoor seating), while review-derived context uniquely captures meal-type and temporal granularity (e.g., brunch, lunch, breakfast) and relational specifics (e.g., husband, work colleagues) largely absent from image context. This limited overlap indicates that image and review context describe substantially different aspects of the same interaction, consistent with the incremental performance gains observed in Table~\ref{tab:context-enhancement}. 

\subsection{Context Category Ablation Analysis}
\label{contextAblation}

Table ~\ref{tab:category-ablation}  reports gating weights and rating-prediction performance across context-category subsets. This ablation analysis was performed on image-based context on its own. When all three categories are active, weights are relatively balanced, indicating that all categories contribute to recommendation performance. When only two categories are available, the gate consistently favors modal context over its paired category, with physical receiving the smallest share. The three-category configuration yields the best recommendation results, consistent with the balanced weighting observed above.

\begin{table}[h]
\caption{Gating weights and rating-prediction performance across context-category subsets. Significance is reported relative to the two-category subsets using a paired two-sided Wilcoxon signed-rank test. *** $p$ < 0.001, ** $p$ < 0.01, * $p$ < 0.05.}
\label{tab:category-ablation}
\renewcommand{\arraystretch}{0.9}
\begin{tabular}{lllll}
\toprule
Category Subset & Physical Weight & Social Weight & Modal Weight & RMSE \\
\midrule
Physical, social, modal & 0.329 & 0.303 & 0.368 & \textbf{0.910***} \\
Physical, social & 0.393 & 0.607 & -- & 0.912 \\
Social, modal & -- & 0.396 & 0.604 & 0.912 \\
Physical, modal & 0.392 & -- & 0.608 & 0.911 \\
\bottomrule
\end{tabular}
\end{table}

\section{Conclusions}
\label{sec:conclusions}
In this work, we propose the use of contextual signals from images to extend established CARS sources such as location, timestamp, and text-derived context. To evaluate the usefulness of this signal, we present ICE-Fuse, a context evaluation pipeline that acquires physical, social, and modal context from review images and integrates it into a recommendation system. We find that image-based context does not outperform established contextual signals as a standalone predictor, but it consistently improves them when combined. A semantic analysis further shows that image- and text-derived context capture largely distinct concepts, supporting the view that image context is a complementary, rather than an overlapping, source of information. Although modest, these improvements are meaningful, as small gains in recommendation accuracy can drive substantial economic value \cite{gomezuribe2016netflix}. 

This study has several limitations. Context quality is assessed indirectly through downstream performance, in the absence of ground-truth labels. Available structured context is limited to city and coarse temporal features; finer-grained signals such as time of day are unavailable. In follow-up work we plan to evaluate image-based context across additional datasets, domains, and algorithms, and to study which visual cues matter most.

\bibliographystyle{ACM-Reference-Format}
\bibliography{references}

\appendix
\section{Prompts}
\label{app:prompts}
\subsection{Physical Context Prompt}

BACKGROUND: You are analyzing an image from a TripAdvisor restaurant review for a context-aware recommender system, focusing on PHYSICAL context. 
            
DEFINITION: Contextual attributes are independent of user and item attributes and can affect ratings. Rating = f(User, Item, Context). Physical context represents the time, position, and activity of the user, as well as the weather, light, and temperature of the dining situation. 
            
SCOPE: Focus on physical-environmental cues visible in the image. Examples of relevant signals: indoor or outdoor seating, natural or artificial lighting, time-of-day cues, seasonal indicators, weather visible through windows, urban or rural surroundings, lighting quality and mood, decor style, crowdedness, overall ambiance of the space.            
            
OUTPUT: Provide only a concise physical context description, in free text. If no meaningful physical context can be determined, respond with $"no\_context"$. Avoid speculation words like "likely", "appears", "suggests", "potentially" and avoid phrases like "I can see", "The image shows", "Context:", "The setting is", "This appears to be". 

\subsection{Social Context Prompt}

BACKGROUND: You are analyzing an image from a TripAdvisor restaurant review for a context-aware recommender system, focusing on SOCIAL context.
                                         
DEFINITION: Contextual attributes are independent of user and item attributes and can affect ratings. Rating = f(User, Item, Context). Social context represents the presence and role of others around the diner, including whether the diner is alone or in a group, and the social configuration of the dining experience.                                         
            
SCOPE: Focus on social-configuration cues visible in the image. Examples of relevant signals: number of people visible, group size and composition, relational configuration (couple, family, friends, business associates), body language, table set for one or two or a group, presence of children, signs of formal or casual interaction. 
                                         
OUTPUT: Provide only a concise social context description, in free text. If no meaningful social context can be determined, respond with $"no\_context"$. Avoid speculation words like "likely", "appears", "suggests", "potentially" and avoid phrases like "I can see", "The image shows", "Context:", "The setting is", "This appears to be". 

\subsection{Modal Context Prompt}

 BACKGROUND: You are analyzing an image from a TripAdvisor restaurant review for a context-aware recommender system, focusing on MODAL context. 
            
DEFINITION: Contextual attributes are independent of user and item attributes and can affect ratings. Rating = f(User, Item, Context). Modal context represents the current state of mind of the diner, the diner's goals, mood, and experience. 
            
SCOPE: Focus on modal cues visible in the image. Examples of relevant signals: purpose or occasion of the visit (relaxation, romantic date, celebration, business meeting, family gathering, exploration); celebratory items such as cake, candles, balloons, or gifts; aesthetic framing suggesting the diner's experiential focus; visual indicators of mood such as calm, lively, romantic, fun, or festive.            
            
OUTPUT: Provide only a concise modal context description, in free text. If no meaningful modal context can be determined, respond with $"no\_context"$. Avoid speculation words like "likely", "appears", "suggests", "potentially" and avoid phrases like "I can see", "The image shows", "Context:", "The setting is", "This appears to be".          

\end{document}